\documentclass[prb,aps,showpacs,prb,amsmath,floatfix,twocolumn]{revtex4}
\usepackage[]{graphicx}
\usepackage{graphicx}
\usepackage{dcolumn}
\usepackage{bm}

\begin{document}
\title{Designer switches: Effect of crystal planes on time-dependent 
electron transport through an interacting quantum dot}
\author{A. Goker$^{1}$, Z. Zhu$^{2}$,  U. Schwingenschlogl$^{2}$}

\affiliation{$^1$
Department of Physics, \\
Bilecik University, 11210, Gulumbe, Bilecik, Turkey
}

\affiliation{$^2$
King Abdullah University of Science and Technology, \\
Physical Sciences and Engineering Division, Thuwal, Saudi Arabia
}

\date{\today}

\begin{abstract}
The time-dependent non-crossing approximation is 
utilized to determine the effects of the crystal 
planes of gold contacts on time dependent current 
through a quantum dot suddenly shifted into the 
Kondo regime via a gate voltage. For an asymmetrically 
coupled system, instantaneous conductance exhibits 
complex fluctuations. We identify the frequencies 
participating in these fluctuations and they turn out 
to be proportional to the separation between the 
sharp features in the density of states and the Fermi 
level in agreement with previous studies. Based on this 
observation, we predict that using different crystal 
planes as electrodes would give rise to drastically 
different transient currents which can be accessed 
with ultrafast pump-probe techniques.     
\end{abstract}



\thispagestyle{headings}

\maketitle
Time-dependent electron transport in single electron 
devices is a subject of fundamental importance in 
molecular electronics since it is widely believed 
that these devices have the potential to replace the 
conventional MOSFET transistors \cite{semiconductor} 
in the near future thanks to the rapid advances in 
nanotechnology. Detection of electrons in real time 
\cite{LuetAl03Nature} is expected to play an important 
role in development of quantum computers 
\cite{ElzermanetAl04Nature} and single electron guns 
\cite{FeveetAl07Science} as well.

Time-dependent current arising from sudden switching 
of the gate or bias voltage \cite{NordlanderetAl99PRL,
PlihaletAl00PRB,MerinoMarston04PRB} has been shown to
exhibit various time scales \cite{PlihaletAl05PRB,
IzmaylovetAl06JPCM}. Moreover, interference
between the Kondo resonance and the sharp features
in the contacts' density of states emerges
in the long timescale associated with the formation
of this many-body resonance \cite{GokeretAl07JPCM}.
Effective one-electron theories indicate that the 
transport properties in steady state depend on 
the the type of electrode metal \cite{KondoetAl09JPCM}, 
the contact-structure \cite{KondoetAl06PRB} and 
the indices of crystal planes of electrode metal 
\cite{Wangetal10JAP}. Same approach also predicts 
that the electrode metals would alter the rectifying 
performance of the device \cite{DengetAl09APL}. 

In previous studies of single electron 
devices, both the Green's function techniques 
\cite{ZhuetAl05PRB,MaciejkoetAl06PRB,GokeretAl07JPCM}
and nonequilibrium diagrammatic Monte Carlo 
method \cite{WerneretAl09PRB,SchmidtetAl08PRB} 
showed that the band structure of the contacts 
has significant influence on the shape of the 
transient current. In all these cases, simple 
unrealistic bands have been used for ease in 
calculations. In this letter, we will carry out 
a comparative study using three different crystal 
planes of gold as contacts to determine the 
transient current through a single electron 
device in the Kondo regime. A unique approach 
which involves using the outcome of an ab-initio 
calculation as an input in a many-body technique 
will be adopted.
  
Single impurity Anderson Hamiltonian representing a 
discrete spin degenerate level of energy $\epsilon_{dot}$ 
connected to Fermi liquid electrodes describes the 
physical behaviour of this system adequately. Upon 
applying the auxiliary boson transformation to the 
Anderson Hamiltonian, it is transformed into
\begin{eqnarray}
H(t)&=&
\sum_{k\alpha\sigma}\left [\epsilon_{k}n_{k\alpha\sigma}+
V_{\alpha}(\varepsilon_{k\alpha},t)c_{k\alpha\sigma}^{\dag}
b^{\dag}f_{\sigma}+{\rm H.c.} \right]+ \nonumber \\
& & \sum_{\sigma}\epsilon_{dot}(t)n_{\sigma},
\end{eqnarray}
where operators $f_{\sigma}^{\dag}(f_{\sigma})$ and 
$c_{k\alpha\sigma}^{\dag}(c_{k\alpha\sigma})$ with 
$\alpha$=L,R create(destroy) an electron with spin 
$\sigma$ within the dot and in the left(L) and right(R) 
contacts respectively. The corresponding number operators
are $n_{\sigma}$ and $n_{k\alpha\sigma}$ whereas the 
hopping amplitude for each contact is denoted with 
$V_{\alpha}$. Operator $b^{\dag}(b)$ creates(destroys) 
a massless boson within the dot and double occupancy 
is prevented by setting the sum of the number of bosons 
and the electrons to unity. If one neglects the explicit 
time dependency of the hopping matrix elements, the 
coupling of the quantum dot to the contacts can be cast as 
$\Gamma_{L(R)}(\epsilon)=\bar{\Gamma}_{L(R)} \xi_{L(R)}(\epsilon)$.
In this expression, $\bar{\Gamma}_{L(R)}$ is a constant 
determined by $\bar{\Gamma}_{L(R)}=2\pi|V_{L(R)}(\epsilon_f)|^2$ 
and $\xi_{L(R)}(\epsilon)$ is the density of states of 
the contacts. 

\begin{figure}[htb]
\centerline{\includegraphics[angle=0,width=6.4cm,height=5.6cm]{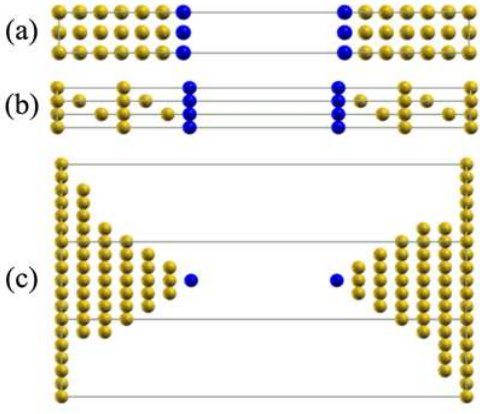}}
\caption{
This figure shows the three geometries, (a) (001)-surface, 
(b) (111)-surface and (c) (111)-pyramide, used to simulate 
the profile of Au electrode. The Au atoms in blue color are 
the atoms relevant to the transport properties between 
electrodes in all three structures.
}
\label{Fig0}
\end{figure}

\begin{figure}[htb]
\begin{center}$
\begin{array}{c}
\includegraphics[angle=0,width=6.8cm,height=5.0cm]{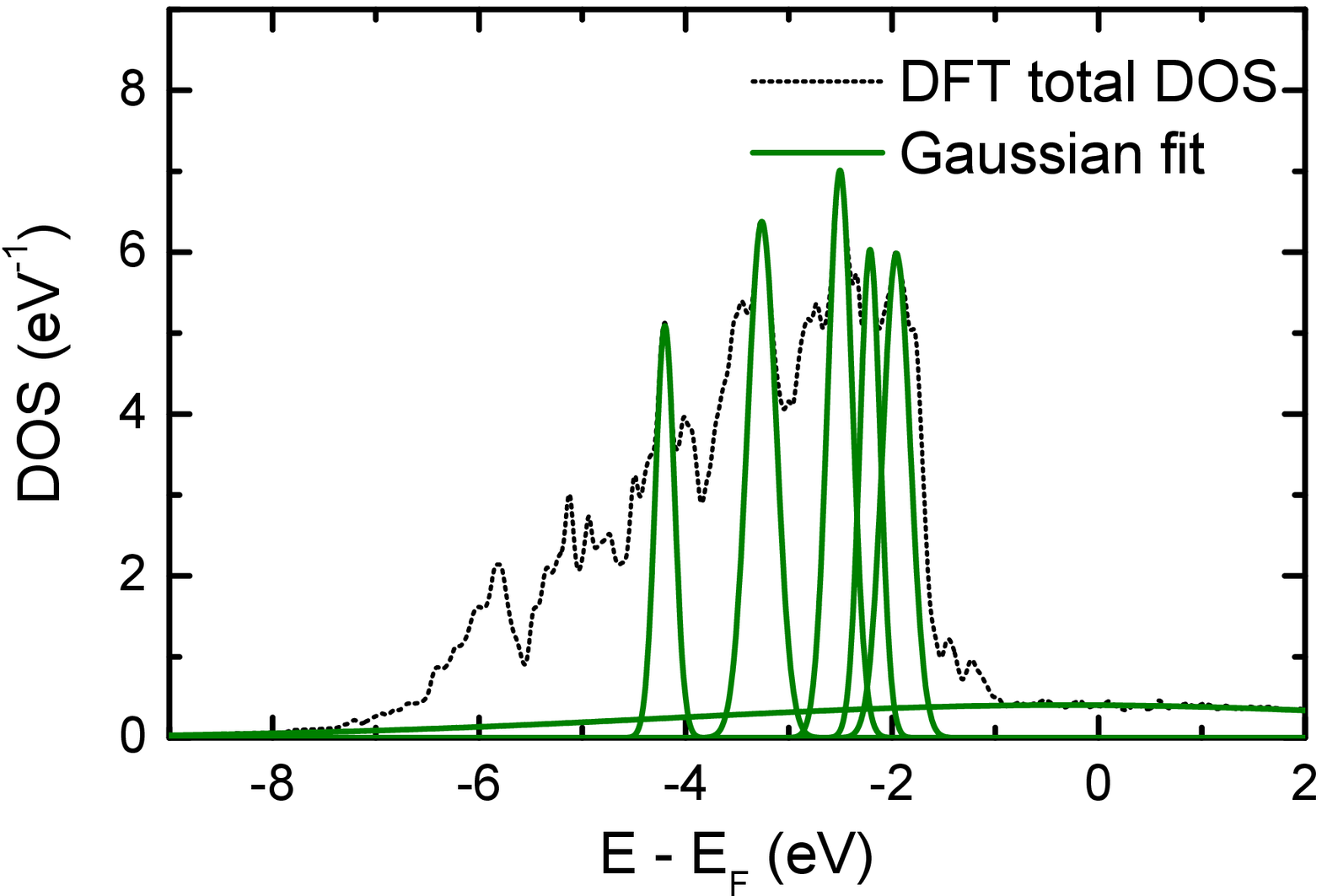} \\
\includegraphics[angle=0,width=6.8cm,height=5.0cm]{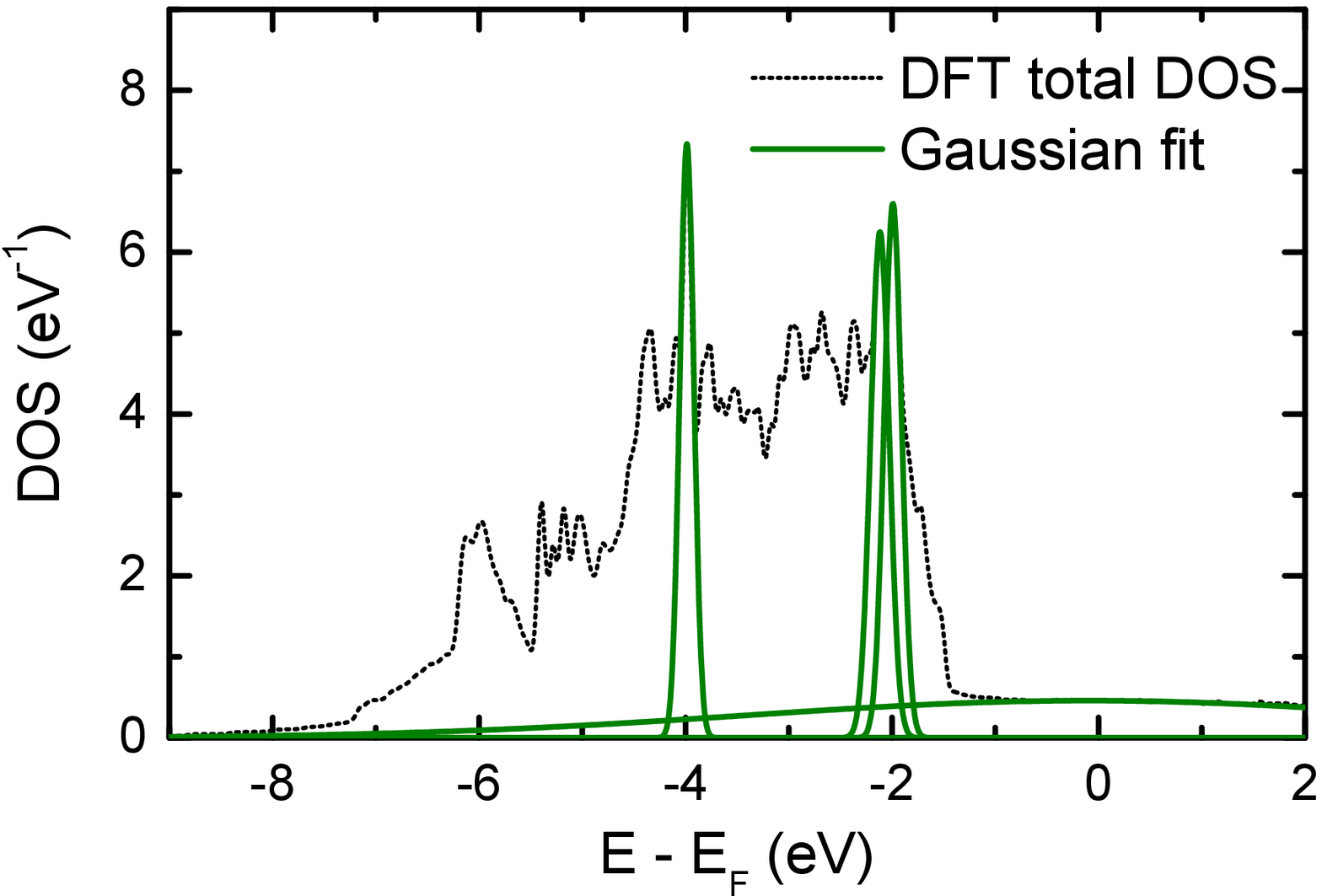} \\
\includegraphics[angle=0,width=6.8cm,height=5.0cm]{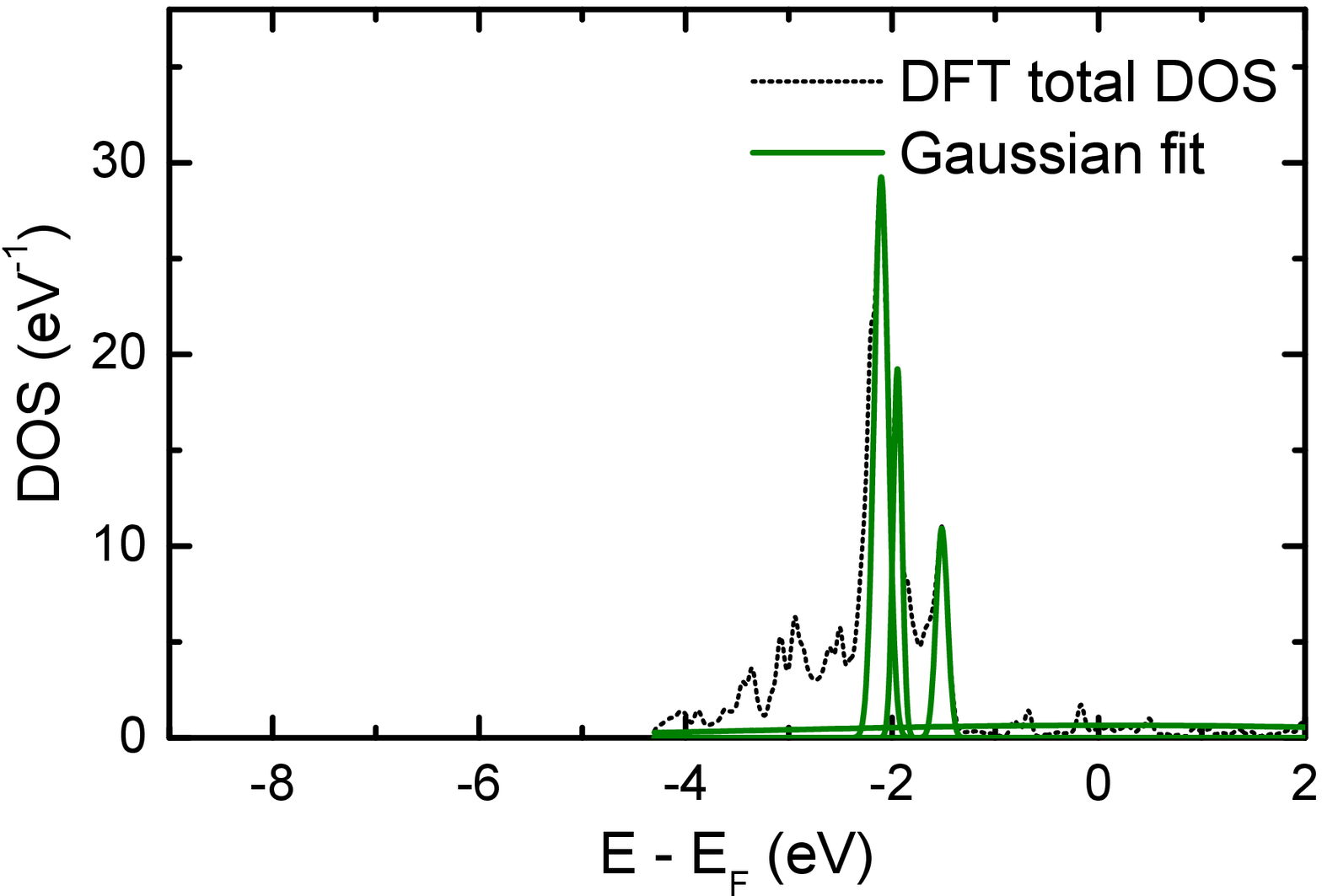}
\end{array}$
\end{center}
\caption{ 
Density of states of (001) surface, (111) surface and (111) 
pyramid calculated using DFT is shown with black dashed curve 
from top to bottom respectively as a function of separation 
from the Fermi level. Each Gaussian used to capture the sharp 
features and the Fermi level is also shown with green curves 
in each geometry.
}
\label{Fig1}
\end{figure}

The density-functional theory (DFT) calculations for the 
density of states (DOS) of the Au contacts have been 
performed with the full-potential linearized augmented 
plane wave (FP-LAPW) method using WIEN2K package 
\cite{Blahaetal01Book}. As shown in Fig.~\ref{Fig0},
three geometries, i.e. (001)-surface, (111)-surface and 
(111)-pyramid, have been adopted to simulate the Au 
electrode profile. In order to build the structures of 
these three geometries, atomic slabs with 13 Au layers 
have been used. The distance between the two opposite 
electrodes is 30 Bohr radius in all three geometries. 
The exchange-correlation potential of the generalized 
gradient approximation within the Perdew, Burke, and 
Ernzerhof (GGA-PBE) form \cite{Perdewetal96PRL} has been 
used in all calculations. The plane wave cut-off has been 
determined by $R_{mt}K_{max}$ = 6.5 and $l_{max}$ = 10. 
K-mesh of 30x30x3, 36x36x2 and 6x6x2 have been adopted 
for (001)-surface, (111)-surface and (111)-pyramid 
respectively. Note that only the DOS of topmost atoms, 
which are relevant to the transport properties between 
electrodes in the surface/pyramid structures, will be 
used in the following many body calculations. The resulting 
density of states is shown in Fig.~\ref{Fig1}.

We performed a fitting procedure involving a linear combination of Gaussians given by 
\begin{equation}
\rho(\epsilon)=\sum_i \frac{\alpha_i}{\zeta_i \sqrt{0.5\pi}}exp(-2(\frac{\epsilon-\epsilon_i}{\zeta_i})^2),
\end{equation}
to the actual DFT data such that all of the sharp peaks 
would be captured. An extra broad Gaussian has been added
in each case to ensure that the density of states at the 
Fermi level is non-zero and the entire bandwidth of the
material is covered. The outcome of this data fitting 
procedure that involves Gaussians of varying linewidth 
and peak position is also shown in Fig.~\ref{Fig1} 
for three different geometries. 
In the remaining part of this letter, we will shift 
to atomic units where $\hbar=k_B=e=1$.




The Dyson equations are solved for the retarded and less 
than Green's functions in a discrete two-dimensional grid. 
This requires the self-energy of electron and massless boson 
as an input. We use non-crossing approximation to define the 
self-energies and close these coupled integro-differential 
equations \cite{ShaoetAl194PRB,IzmaylovetAl06JPCM}.  
The net current follows from the resulting Green's functions 
denoted by $G_{pseu}^{<(R)}(t,t')$ and $B^{<(R)}(t,t')$. The 
general expression for the net current $I(t)$ \cite{JauhoetAl94PRB} 
can be recast by using the slave boson decomposition method 
\cite{GokeretAl07JPCM}. The equation we will use in this 
letter to determine the net current is then written as
\begin{eqnarray}
& & I(t) =-2(\bar{\Gamma}_{L}-\bar{\Gamma}_{R})\textit{Re} \left (\int_{-\infty}^{t} dt_1
\xi_{o}(t,t_1)h(t-t_1)\right)+\nonumber \\
& & 2\bar{\Gamma}_{L} Re \left (\int_{-\infty}^{t} dt_1 
(\xi_{o}(t,t_1)+\xi_{u}(t,t_1)) f_{L}(t-t_1) \right)- \nonumber \\
& & 2\bar{\Gamma}_{R} Re \left (\int_{-\infty}^{t} dt_1
(\xi_{o}(t,t_1)+\xi_{u}(t,t_1)) f_{R}(t-t_1)\right)\nonumber \\
\label{final}
\end{eqnarray}
where $\xi_{o}(t,t_1)=G_{pseu}^{<}(t,t_1)B^{R}(t_1,t)$ and 
$\xi_{u}(t,t_1)=G_{pseu}^{R}(t,t_1)B^{<}(t_1,t)$. In Eq.~(\ref{final}),
$f_L(t-t_1)$ and $f_R(t-t_1)$ are the convolution of $\xi(\epsilon)$ 
with the Fermi-Dirac distributions of contacts and $h(t-t_1)$ is the 
Fourier transform of $\xi(\epsilon)$ \cite{GokeretAl07JPCM}. The 
conductance $G$ is equal to the current divided by the bias voltage 
$V$. In the following discussion $\eta=\frac{\bar{\Gamma}_{L}}{\bar{\Gamma}_{tot}}$, 
where $\bar{\Gamma}_{tot}=\bar{\Gamma}_{L}+\bar{\Gamma}_{R}$, 
will be referred as the asymetry factor.

Kondo effect is a many-body resonance pinned to 
the Fermi levels of the contacts in the dot density of 
states and manifests itself as an enhancement in the 
conductance at low temperatures. The linewidth of the 
Kondo resonance is characterized by an energy scale $T_K$ 
(Kondo temperature) given by
\begin{equation}
T_K \approx \left(\frac{D\Gamma_{tot}}{4}\right)^\frac{1}{2}
\exp\left(-\frac{\pi|\epsilon_{\rm dot}|}{\Gamma_{tot}}\right),
\label{tkondo}
\end{equation}
where $D$ is the half bandwidth of the conduction electrons 
and $\Gamma_{tot}=\bar{\Gamma} \xi(\epsilon_f)$.

\begin{figure}[htb]
\begin{center}$
\begin{array}{c}
\includegraphics[angle=0,width=6.8cm,height=5.0cm]{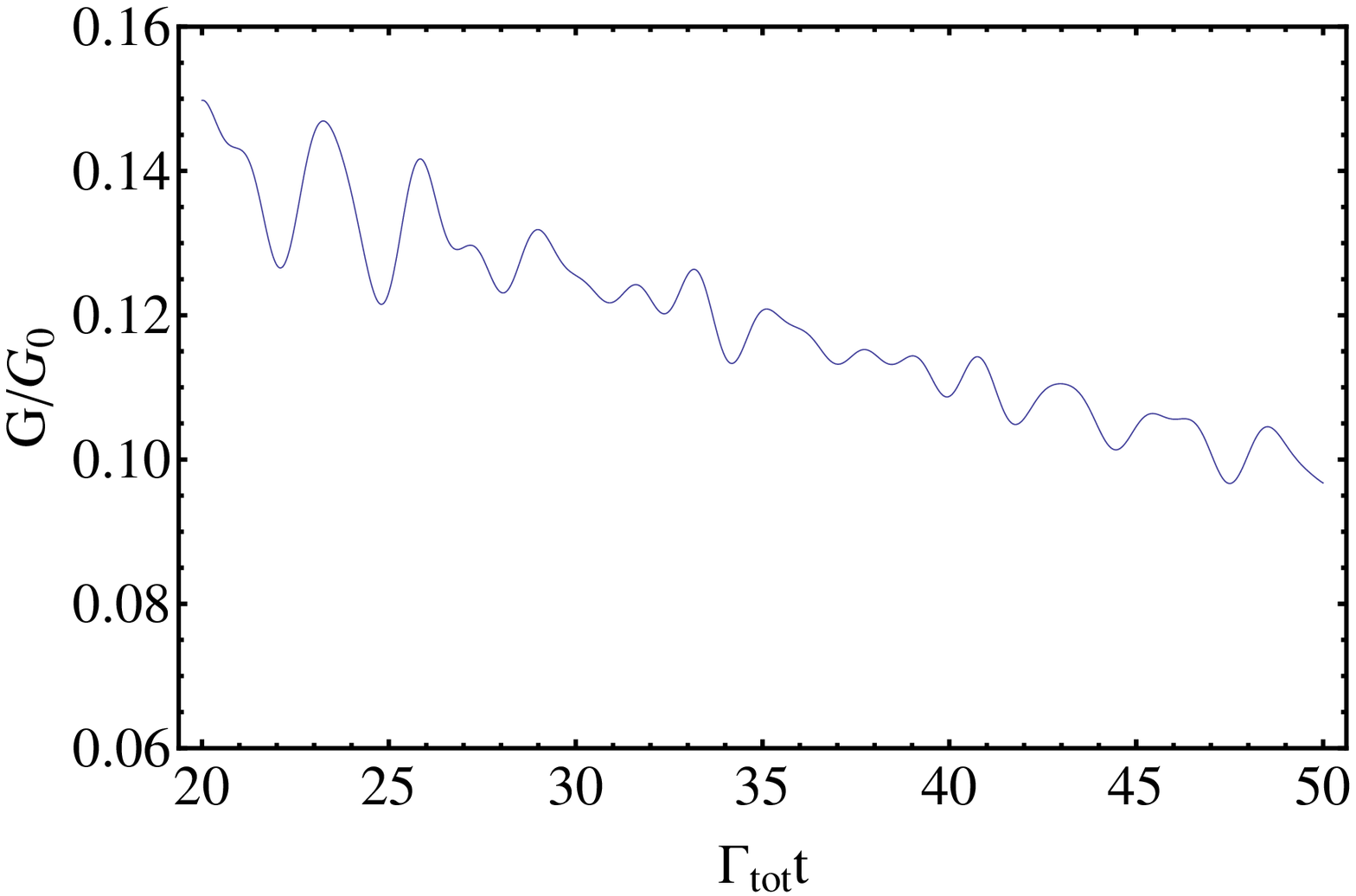} \\
\includegraphics[angle=0,width=6.8cm,height=5.0cm]{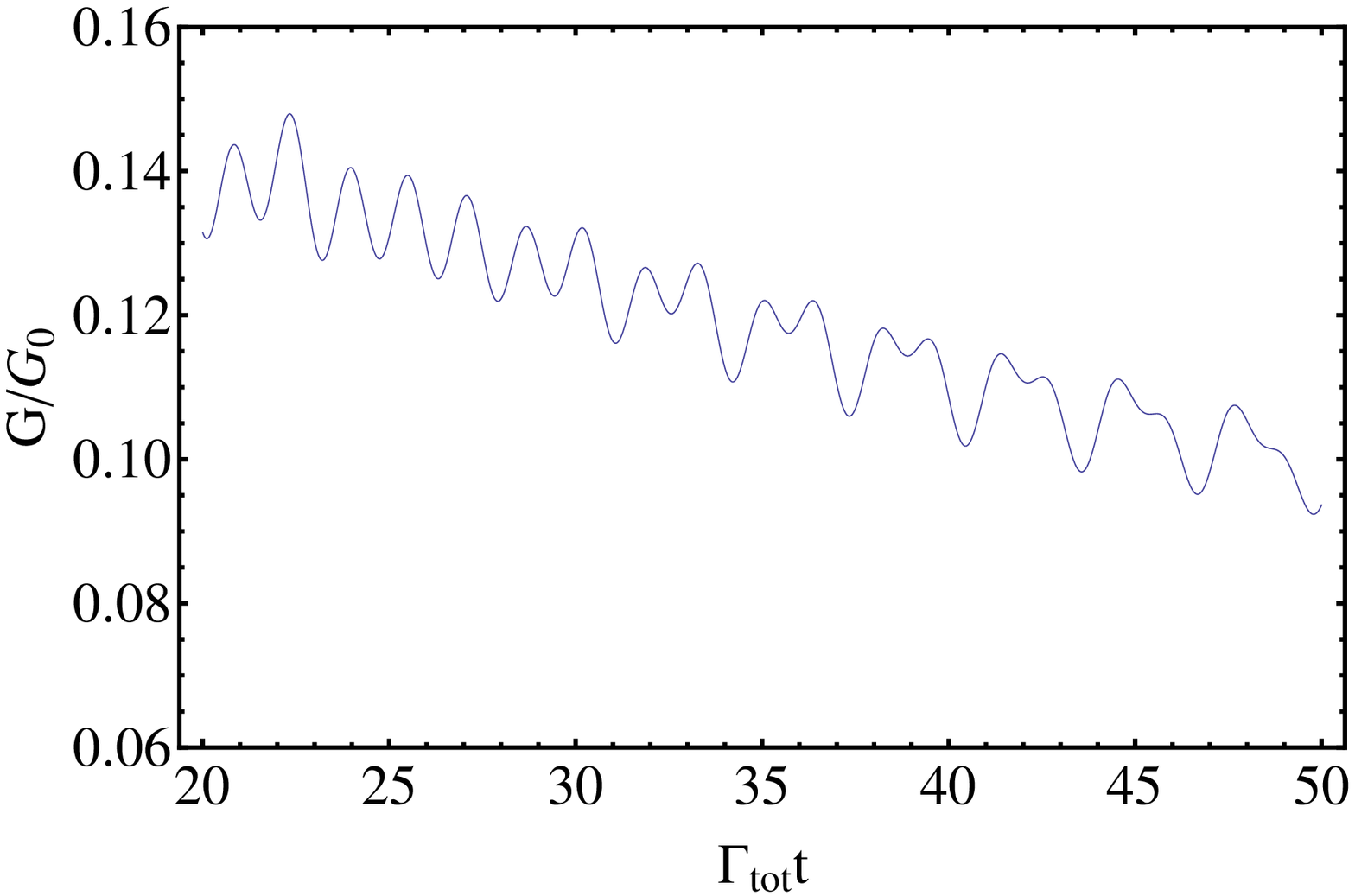} \\
\includegraphics[angle=0,width=6.8cm,height=5.0cm]{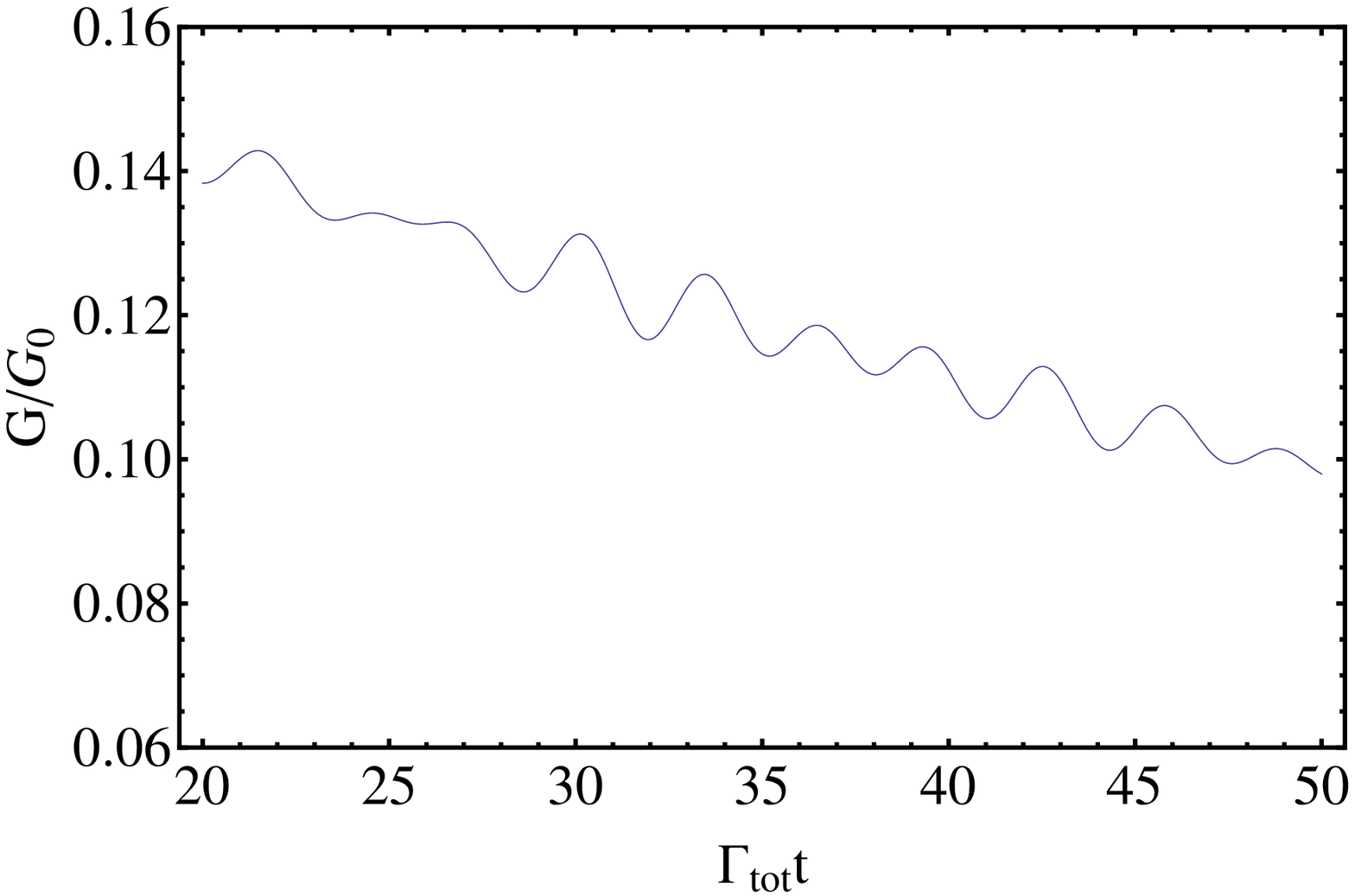}
\end{array}$
\end{center}
\caption{
Panels from top to bottom show the instantaneous conductance versus time in 
Kondo timescale after the dot level has been switched to its final position for 
(001) surface, (111) surface and (111) pyramid respectively with an asymmetry 
factor of 0.9 at T=0.009$\Gamma_{tot}$ in infinitesimal bias.
}
\label{Fig2}
\end{figure}

We will investigate the transient current for the case 
where the dot level is switched from $\epsilon_1=-4\Gamma_{tot}$ 
to $\epsilon_2=-2\Gamma_{tot}$ at t=0 with a gate voltage. 
For all three geometries, a transition from a non-Kondo
state to a Kondo state takes place. Note that the Kondo temperature
in the final state would be slightly lower for (111) pyramide
than the other two geometries as a result of shorter 
conduction electron bandwidth as seen in Fig.~\ref{Fig1}.

Transient current in Kondo timescale is depicted for all 
geometries in Fig.~\ref{Fig2} after switching to the final
dot level in infinitesimal bias. We do not display the
short timescale where where the conductance reaches a 
maximum before it starts to fall off for large asymmetry 
factors due to striking similarity with previous studies 
\cite{GokeretAl07JPCM}. The most remarkable feature is the 
drastic difference in conductance fluctuations. This effect 
stems purely from the difference in band structure of the 
contacts since all other parameters are held constant.
We should point out that the slight difference between the
Kondo temperature of (111) pyramide and the other two
geometries does not change the relative behaviour
of the fluctuations for these geometries because
$T/T_K$ is known to alter only the amplitude of the
fluctuations \cite{GokeretAl07JPCM}, not the overall
pattern of the transient current.  
 
It is clear that these fluctuations are the result
of an admixture of sinusoidal oscillations with
different frequencies and amplitudes. Each individual
frequency can be extracted by taking the Fourier 
transform of the current. We found that the 
frequencies are proportional to the separation 
between the peak positions and the Fermi level 
for all cases. This explains why transient current
for (001) surface exhibits a more erratic pattern 
compared to the others as five distinct frequencies
are involved in it. Moreover, there are other
peaks that appear in actual DFT data but we did 
not include them in our fitting since they have 
negligible contribution to the fluctuation pattern. 
The oscillation amplitude associated with them is 
probably too small to have a discernible effect 
because either they are located far away from the 
Fermi level or the peaks are not prominent compared 
to the surrounding structure.

The results presented here describes a way to probe
the detailed band structure of a molecular switch
by measuring the transient current flowing through it.
It is possible to capture the transient current
fluctuations occurring in the femtosecond timescale
with ultrafast pump-probe techniques \cite{Teradaetal10JPCM}.
This would pave the way for designing custom switches
for future organic computers.    
 


\bibliographystyle{iopams}
\bibliography{gen}

\end{document}